\documentclass{article}
\usepackage{amsmath,amsthm}
\usepackage{mathtools}
\mathtoolsset{showonlyrefs}

\usepackage{ifxetex,ifluatex}
\ifxetex
\usepackage{xltxtra}
\else
\ifluatex
\usepackage{luatextra}
\else
\usepackage{amssymb}
\usepackage[utf8]{inputenc}
\usepackage[T1]{fontenc}
\usepackage{mathpazo,tgcursor,tgtermes}
\usepackage[scale=.9]{tgheros}
\usepackage{textcomp}
\csname else\expandafter\endcsname
\romannumeral -`0
\fi
\fi
\iftrue\relax%
\defaultfontfeatures{Ligatures=TeX}
\usepackage[math-style=TeX, vargreek-shape=TeX]{unicode-math}
\AtBeginDocument{%
}
\fi

\usepackage[footnotesize,bf]{caption}

\usepackage[letterpaper,margin=1in]{geometry}
\usepackage{authblk}

\usepackage[authoryear]{natbib}
\usepackage{color}

\usepackage{algorithm}
\usepackage{algorithmic}

\usepackage{enumitem}
\setenumerate[1]{label=(\roman*)}
\usepackage{xparse}

\usepackage[hyphens]{url}
\usepackage[bookmarksnumbered,bookmarksopen,unicode]{hyperref}
\usepackage{doi}

\newlist{enumerate*}{enumerate*}{1}
\setlist[enumerate*]{label=(\arabic*),
  itemjoin={{, }}, itemjoin*={{, and }}, after={.}}

\newcommand{\R}{\mathbb{R}}

\DeclarePairedDelimiter{\abs}{\lvert}{\rvert}

\DeclareMathOperator{\tr}{Tr}

\DeclareMathOperator{\KLdivergence}{KL}

\DeclareMathOperator*{\argmin}{arg\,min}

\DeclareMathOperator{\diag}{diag}

\newcommand*{\Algorithm}[1]{\textsc{#1}}

\newcommand*{\set}[2]{\left\{#1\,\middle|\,#2\right\}}

\DeclarePairedDelimiterX\inp[2]{\langle}{\rangle}{#1,#2}

\NewDocumentCommand{\probability}{d()om}{%
  \operatorname{\mathbb{P}}%
  \IfValueT{#1}{\sb{#1}}%
  \left[#3%
    \IfValueT{#2}{\,\middle|\,#2}\right]}
\NewDocumentCommand{\expectation}{d()om}%
  {\operatorname{\mathbb{E}}%
    \IfValueT{#1}{\sb{#1}}%
    \left[#3%
      \IfValueT{#2}{\,\middle|\,#2}\right]}
\NewDocumentCommand{\variance}{d()om}%
  {\operatorname{Var}%
    \IfValueT{#1}{\sb{#1}}%
    \left[#3%
      \IfValueT{#2}{\,\middle|\,#2}\right]}

\NewDocumentCommand{\norm}{osm}{%
  \IfBooleanT{#2}{\left}\lVert
    #3%
    \IfBooleanT{#2}{\right}\rVert
  \IfValueT{#1}{\sb{#1}}}

\newtheorem{theorem}{Theorem}[section]
\newtheorem{lemma}[theorem]{Lemma}
\newtheorem{proposition}[theorem]{Proposition}

\theoremstyle{remark}
\newtheorem{remark}[theorem]{Remark}

\theoremstyle{definition}

\hypersetup{
  pdftitle={An efficient high-probability algorithm
    for linear bandits},
  pdfauthor={Gábor Braun, Sebastian Pokutta},
  pdfkeywords={regret minimization, bandit feedback,
    implicit exploration},
  pdfsubject={MSC2000 Primary 68Q32; 
    Secondary 68Q25 
  }}

\title{An efficient high-probability algorithm \\ for Linear Bandits}
\date{October 13, 2016}
\author[1]{Gábor Braun}
\affil[1]{ISyE, Georgia Institute of Technology,
  Atlanta, GA,
  USA.
  \textit{Email:}~gabor.braun@isye.gatech.edu}

\author[2]{Sebastian Pokutta}
\affil[2]{ISyE, Georgia Institute of Technology,
  Atlanta, GA,
  USA.
  \textit{Email:}~sebastian.pokutta@isye.gatech.edu}

\begin{document}
\maketitle{}
\begin{abstract}
  For the linear bandit problem,
  we extend the analysis of algorithm \Algorithm{CombEXP} from
  \cite{CombEXP2015} to the \emph{high-probability} case against
  \emph{adaptive} adversaries, allowing actions to come from an
  \emph{arbitrary polytope}.
  We prove
  a high-probability regret of \(O(T^{2/3})\)
  for time horizon \(T\).
  While this bound is weaker than the optimal \(O(\sqrt{T})\) bound
  achieved by
  \Algorithm{GeometricHedge} in \cite{RegretLinearBandit2008},
  \Algorithm{CombEXP} is computationally
  \emph{efficient}, requiring only an efficient linear optimization
  oracle over the convex hull of the actions.
\end{abstract}

\section{Introduction}
\label{sec:introduction}

We study sequential prediction problems with
linear losses and bandit feedback against an adaptive adversary.
At every round \(t\) the forecaster chooses an action \(x_{t}\),
and the adversary chooses a loss function \(L_{t}\),
and the forecaster suffers the loss \(L_{t} (x_{t})\).
The forecaster learns only the suffered loss after each round,
while the adversary learns the forecaster's action \(x_{t}\).
The forecaster's aim is to minimize
\emph{regret}, which is the difference between the incurred loss and
the loss of the best
single action in hindsight:
\[\sum_{t\in T} L_{t}(x_{t}) - \min_{x \in A} \sum_{t\in T} L_{t}(x).\]
In this work we focus on establishing regret bounds holding with
high-probability with an efficient algorithm.

For algorithms with bandit feedback, \emph{exploration} (occasionally
playing random actions for learning) is a crucial feature, however it
does not have to be explicit as recently shown in
\cite{ExploreNoMore2015}, where exploration is achieved
via skewing loss estimators.
One of the most studied regret
minimization algorithm is \Algorithm{EXP}, which iteratively updates
the probabilities of each action via multiplication with factors
exponential in its (estimated) loss.  The variant \Algorithm{EXP3} for
multi-armed bandit problems first appeared in \cite{EXP3-2002},
however optimal high-probability regret bounds were first achieved in
\cite{BeatAdaptive2006}.  The linear bandit setting is a
generalization of the multi-armed bandit setting where, utilizing the
linearity of losses, the goal is to improve the dependence on the
number of actions in the regret bound, which might be exponential in
the dimension \(n\).
At the same time linear losses come naturally into play
when considering actions
with a combinatorial structure, such as e.g., matchings, spanning
trees, \(m\)-sets;
see \cite{ComBandit2012,audibert2013regret} for an extensive
discussion. For the linear bandit setting, the \Algorithm{EXP}-variant
\Algorithm{ComBand} (Combinatorial Bandit)
from \cite{ComBandit2012} has optimal
\(O(\sqrt{T})\)
expected regret, and in \cite{RegretLinearBandit2008}
the modified version \Algorithm{GeometricHedge}
achieves \(O(\sqrt{T})\) regret with high probability.
While these regret bounds practically do not depend
on the number of actions, both
maintain a distribution over the (possibly exponentially large) action
set \(A\),
which is infeasible in general due to the large data size, even though
\Algorithm{ComBand} is still efficient for many specific problems.
Recently, a modification of the \Algorithm{ComBand} algorithm
called \Algorithm{CombEXP} (see Algorithm~\ref{alg:CombEXP}) was
derived in \cite{CombEXP2015}, which achieves general computational
efficiency by \emph{not} maintaining a distribution of \(x_{t}\),
but only the \emph{desired expectation} \(\hat{x}_{t}\)
of the distribution, and generating a new sparse approximate
distribution at every round.

In this work we provide a \emph{high-probability} regret bound of
\(O(T^{2/3})\) for \Algorithm{CombEXP} against \emph{adaptive}
adversaries, while generalizing it to general polytopes.
The obtained bounds are \emph{any-time}, i.e., the parameter choice is
independent of the time horizon \(T\).
Finally, our algorithm maintains computational efficiency given
an efficient linear programming oracle over the underlying polytope
(the convex hull of actions).
For comparison,
we also show an \(O(T^{2/3})\)
regret in the high-probability setting for the original
\Algorithm{ComBand}.

The maximal matching problem is a good example
where the linear programming oracle approach is useful,
as it has a polynomial time linear optimization algorithm
\cite{edmonds1965maximum},
but no polynomial-size polyhedral description
\cite{rothvoss2013matching}.

\subsection*{Related work}
\label{sec:related-work}

Our work is most closely related to the line of works on
combinatorial bandit problems.
The algorithm \Algorithm{ComBand}
first appeared in \cite{ComBandit2012},
while \Algorithm{GeometricHedge}
comes from \cite{RegretLinearBandit2008},
and \Algorithm{CombEXP} appeared in \cite{CombEXP2015}.
Using interior point methods,
an efficient algorithm with \(O(\sqrt{T})\) expected regret
for linear bandit problems
has been established in \cite{EfficientLinearBandit2008}.

For multiarmed bandit problems, the original version of
\Algorithm{EXP3} has high-probability regret \(\Omega(T^{2/3})\)
against some adaptive adversaries
\cite[Theorem~1.2]{BeatAdaptive2006}, however variants with optimal
\(O(\sqrt{T})\)
regret exists, e.g., using accountants to control the exploration rate
(see \cite{BeatAdaptive2006}), or via the recent \Algorithm{EXP3-IX}
with implicit exploration (see \cite{ExploreNoMore2015}).

For convex loss functions,
optimal high-probability regret bounds have been obtained in
\cite{OptimalBCO2016} with running time being poly-exponential in the
dimension, and in \cite[Theorem~1]{KernelConvexBandit2016} with
polynomial running time provided the number of constraints of the
underlying polytope is polynomial in the dimension.
Optimal regret bounds in expectation was first obtained in
\cite{ConvexBandit2015}.
However the case of convex loss does not subsume the
combinatorial/linear case, as with convex loss \emph{all inner points} of the
convex set are actions;
with linear losses the actions are limited to the
vertices of the underlying polytope
in most cases.

We refer the interested reader to the excellent survey of
\cite{StochasticMultiBandit2012} on bandit problems.

\subsection*{Contribution}
\label{sec:contribution}

Our main contribution is a \emph{high-probability} regret bound for
\Algorithm{CombEXP} from \cite{CombEXP2015} for \emph{adaptive
adversaries} over actions coming from \emph{arbitrary polytopes} \(P \subseteq
\R^n\). Our algorithm, being a slight generalization of
\Algorithm{CombExp}, maintains computational efficiency.  In
particular, our contribution can be summarized as follows: 

\begin{enumerate}
\item \emph{High-probability bounds for an efficient algorithm.}
  For \Algorithm{CombEXP} we establish
  a high-probability regret of
\[O\left( 
  \frac{B^2 + n B}{\min\{\lambda, 1\}}
  \cdot \ln \frac{2n + 2}{\delta}
  \right) T^{2/3},
\]
  with probability \(1-\delta\),
  where \(B\) is the \(\ell_{2}\)-diameter of \(P\),
  and \(\lambda\) is a lower bound on the smallest
  eigenvalue of the exploration covariance matrix,
  see Theorem~\ref{thm:CombEXP} for the
  exact regret bound.

  For comparison we show that the same method already provides a
  high-probability regret bound of \(O(T^{2/3})\) for the original
\Algorithm{ComBand}, albeit a suboptimal one as
\Algorithm{GeometricHedge} achieves \(O(\sqrt{T})\) regret. 

\item \emph{Generalization of CombEXP and computational efficiency.} We generalize
  \Algorithm{CombEXP} to actions arising from \emph{arbitrary} polytopes contained
  in \(\R^n\) and to the case of \emph{adaptive adversaries}. We maintain
  computational efficiency of \Algorithm{CombExp}
  providing running times relative to a linear
  programming oracle over the underlying polytope \(P\),
  separating the
  complexity for learning from the complexity of
  linear optimization over \(P\).
\end{enumerate}

All our bounds are \emph{any-time}, i.e., holding uniformly for all
times \(T\). In particular, our parameter choices are independent of
\(T\). 

\subsection*{Outline}
\label{sec:outline}

After a brief summary of the regret minimization framework in
Section~\ref{sec:preliminaries}, we reanalyze \Algorithm{CombEXP} in
Section~\ref{sec:regret-bound-CombEXP}.  For completeness we present a
similar analysis for \Algorithm{ComBand} in
Section~\ref{sec:regret-bound-ComBandit}. 

We relegated various related materials to the the Appendix. In Section~\ref{sec:time-exp3-proj}
we provide an \emph{any-time} version of \Algorithm{EXP} with time-varying parameters
maintaining generalized distributions, defined by an arbitrary convex set in the positive orthant,
instead of the probability simplex. We prove an \(O(\sqrt{T})\) regret bound in the full information case
by standard arguments, which forms the basis for our regret bounds for
the bandit case. In Section~\ref{sec:concentration} we recall concentration inequalities
that we use to establish high-probability bounds.
Finally, in Sections~\ref{sec:projection-KL} and \ref{sec:linear-decomposition}
we provide (already known) efficient algorithms
for projection and distribution generation, which are key components in our algorithms.
We include those for completeness of exposition and to make parameters explicit.

\section{Preliminaries}
\label{sec:preliminaries}

We will briefly recall the regret minimization framework to define our
notation.
In the \emph{sequential prediction problem}
with \emph{linear} losses,
at every round \(t\) the forecaster chooses
an \emph{action} \(x_{t}\) from a finite set \(A \subseteq \R^n\)
and the adversary
chooses a \emph{loss vector} \(L_{t} \in \R^{n}\).
The forecaster suffers
the loss \(\ell_{t} \coloneqq L_{t}^\intercal x_{t}\).
The goal of the forecaster is to minimize the regret
\begin{equation*}
  \sum_{t=1}^{T} L_{t}^{\intercal} x_{t}
  -
  \min_{x \in A}\sum_{t=1}^{T} L_{t}^{\intercal} x
  .
\end{equation*}
Against an oblivious adversary,
who chooses the \(L_{t}\) independently of the forecaster's actions,
this is the extra loss suffered by not
playing the best single action in hindsight.
However, this interpretation is clearly incorrect against an adaptive
adversary (the notion of \emph{policy regret} from
\cite{policyRegret2012} matches this interpretation).
Nevertheless the above notion of regret proved to be useful in many
areas.

With \emph{bandit} feedback the forecaster learns
only the loss \(\ell_{t}\) but not the
actual loss vector \(L_t\).
An \emph{adaptive} adversary learns the forecaster's action \(x_{t}\)
after round \(t\), and can use it in later rounds
to choose his actions.

We make various standard assumptions to bound the regret.
The most important one is that the per round loss is bounded, i.e.,
\(\abs{L_{t}^{\intercal} x} \leq 1\) for all \(x \in A\).
Under reasonably assumptions, this also implies that
the set \(A\) of possible actions \(A\) is bounded and we assume that 
\(\norm[2]{x} \leq B\) and \(\norm[1]{x} \leq B_{1}\),
with suitable positive numbers \(B\), \(B_{1}\).
Clearly, one can always choose \(B_{1} = n B\),
however we obtain finer bounds by keeping them separate.
The bounds \(B_1\) and \(B\) also serve as
a proxy for the sparsity of the \emph{actions}.

Following \cite{ComBandit2012} for \Algorithm{ComBand},
we shall use a fixed
arbitrary distribution \(\mu\) on \(A\)
for exploration, whose fitness for exploration is measured by
a positive lower bound \(\lambda\)
on the smallest eigenvalue of its covariance matrix \(J\):
\begin{equation*}
  J \coloneqq \expectation(y \sim \mu){y y^{\intercal}} \succeq
  \lambda I
  .
\end{equation*}
Here and below we denote by \(M \preceq N\)
that \(N - M\) is a positive semi-definite matrix
for symmetric matrices \(M\) and \(N\).
When \(A\) is small then
\(\mu\) is typically the uniform distribution over \(A\).
For large \(A\), common choices are the uniform distribution on
a barycentric spanner of \(A\) (see \cite{Spanners2014}),
or the distribution on contact points
of the maximal volume ellipsoid contained in the convex hull
\(P\) of \(A\)
arising from John's decomposition
(John's exploration; see \cite{BanditPrice2008}),
transferred to \(A\).
In the latter two cases, \(J = I\) and \(\lambda = 1/n\)
using the scalar product on \(\R^{n}\)
\emph{induced by} the additional structure.
John's ellipsoid can be approximately estimated with
a worse lower bound \(\lambda = 1 / n^{3/2}\) by
\cite{GrotschelLovaszSchrijverBook},
however a constant factor approximation is NP-hard
by \cite{ConicProgram2006}.
Recall that a \emph{barycentric spanner} is a linear basis
\(v_{1}, \dotsc, v_{n}\) in \(P\) (the convex hull of \(A\)),
such that every element of \(P\) is a linear combination of the
\(v_{i}\) with coefficients from \([-1, +1]\).
The basis \(v_{1}, \dotsc, v_{n}\) is a
\emph{\(C\)-approximate barycentric spanner} for some \(C > 1\)
if every element of \(P\) is a linear combination of the
\(v_{i}\) with coefficients from \([-C, +C]\).
A \(C\)-approximate barycentric spanners can be efficiently computed
by \(O(n^{2} \ln n / \ln C)\) calls to a linear optimization
oracle over \(P\) by \cite{Routing2004}, which actually computes
a spanner consisting of vertices of \(P\).
In this paper we deliberately avoid using the scalar product induced
by the structure to be able to directly use the bounds available in
the original space of the problem. Fortunately, the uniform distribution
on an approximate barycentric spanner has a close to optimal minimal
eigenvalue \emph{even in the original space}, see
Lemma~\ref{lem:fitness-spanner}, which allows us to preserve sparsity
of the original space. As such we assume that we have access
to an exploration distribution over actions with sparse support of size \(n\),
where \(n\) is the dimension of the vector space, from which we can
efficiently sample. Note that for specific problems exploration
distributions with better minimal eigenvalue can be explicitly
given; we refer the interested reader to \cite{ComBandit2012} and
follow-up work for a large set of such examples. 

Let \(u \coloneqq \expectation(y \sim \mu){y}\)
denote the expectation of \(\mu\)
and let \(e\)
denote the Euler constant.  Instead of dealing directly with \(A\),
it will be more convenient to use the convex hull \(P\)
of \(A\),
then \(A\)
contains the vertex set of \(P\)
(and in many applications the two are equal).  We shall use
the Kullback–Leibler divergence as Bregman divergence of the function
\(f(x_{1}, \dots, x_{n}) = \sum_{i=1}^{n} x_{i} \ln x_{i}\)
for projection:
\begin{equation*}
  \KLdivergence(x, y) = \sum_{i=1}^{n} x_{i} \ln \frac{x_{i}}{y_{i}}
  - \sum_{i=1}^{n} x_{i} + \sum_{i=1}^{n} y_{i}
  .
\end{equation*}

In the following, for a vector \(a \in \R^{n}\)
we will use \(\R^{n}_{>a} \coloneqq \set{x \in \R^{n}}
{x_{i} > a_{i} \text{ for all } i \in [n]}\) to denote the
\emph{\(a\)-positive orthant}. Moreover, a \emph{linear optimization
  oracle (or LP oracle)} over a polytope \(P \subseteq \R^n\)
finds for any linear objective \(c \in \R^{n}\)
a vertex \(x\) of \(P\) minimizing \(c^{\intercal} x\).

In all our bounds below,
the \(O\)-notation only hides an absolute constant,
i.e., all parameters of the algorithms are explicit.
However, in Section~\ref{sec:introduction}
the \(O\)-notation hides also other parameters,
like the dimension \(n\).

\section{A high-probability regret bound for CombEXP}
\label{sec:regret-bound-CombEXP}

We provide an adaptation of \Algorithm{CombEXP}
(Algorithm~\ref{alg:CombEXP}) with an \(O(T^{2/3})\)
regret with high probability against adaptive adversaries, while
maintaining computational efficiency.  In a nutshell, \Algorithm{EXP}
is run on the coordinates of the \emph{desired expectation}
\(\hat{x}_{t}\)
of \(x_{t}\),
and a new distribution over vertices \(x_{t}\)
of \(P\)
is generated in every round.  In order to obtain an efficient
algorithm, we allow errors in the most resource-consuming components
of the algorithm: the projection step and the distribution
generation. The accuracy of distribution generation is controlled by a
parameter \(\varepsilon\),
and helps maintaining a distribution with \emph{sparse} support, to
allow fast sampling and fast computation of the covariance matrix
\(C_{t}\). The positive parameters \(\eta_{t}\), \(\gamma_{t}\)
control the \emph{learning rate} and \emph{exploration rate} of the
algorithm.  The role of the shifting vector \(a \in \R^{n}\)
is to avoid singularity issues with Kullback–Leibler divergence.
Except for the shifting vector \(a\),
these ideas already appeared in \cite{CombEXP2015}.

The algorithm contains four resource-consuming steps:
\begin{enumerate*}
\item\label{item:projection}
  projection (Line~\ref{line:projection})
\item\label{item:distribution}
  distribution generation (Line~\ref{line:distribution})
\item\label{item:sample}
  sampling from the distribution
\item\label{item:covariance}
  computing the covariance matrix
\end{enumerate*}
All the other steps are fast, depending only polynomially on the
dimension.

The major factor for the running time of sampling from the
distribution \ref{item:sample}, and computing the covariance matrix
\ref{item:covariance} is the sparsity of the generated distribution,
i.e., the number of possible outcomes.
Sparse distributions (number of outcomes polynomial in the dimension)
of sufficient accuracy can be efficiently
generated by the decomposition algorithm from \cite{Caratheodory2015},
which we summarize as Algorithm~\ref{alg:linear-decomposition} in
Section~\ref{sec:linear-decomposition} for the reader's convenience.
Common choices of the exploration distribution \(\mu\) are sparse,
as discussed above,
notwithstanding
non-sparse distributions for \(\mu\) are also acceptable
which have an efficient sampling method
and a precomputed covariance matrix.
Therefore we will disregard the complexity of
sampling and computation of the covariance matrix.

Finally, the projection step
(Line~\ref{line:projection}) can be efficiently accomplished by the
Frank–Wolfe algorithm (also called conditional gradient), which we
recall in Algorithm~\ref{alg:projection-KL} in
Section~\ref{sec:projection-KL}.  Note that if
Algorithm~\ref{alg:projection-KL} is used for the projection step, it
already provides a sparse linear decomposition of
the desired expectation \(\hat{x}_{t+1}\)
with accuracy \(\varepsilon = 0\),
and therefore makes a separate linear decomposition step unnecessary.
Nevertheless it might be advantageous for specific polytopes to use a
specialized, more efficient projection algorithm and/or decomposition
algorithm.

All in all, we measure complexity of only the most time-consuming
tasks: projection and linear decomposition,
requiring the linear decomposition to be sparse.
We report complexity of
Algorithms~\ref{alg:projection-KL} and~\ref{alg:linear-decomposition}
mentioned above
in the \emph{total} number of
linear optimization oracle calls over \(P\).
This relative complexity is often useful in applications
where fast linear programming oracles are available.

\begin{algorithm}
  \caption{\Algorithm{CombEXP}}
  \label{alg:CombEXP}
  \begin{algorithmic}[1]
    \REQUIRE polytope \(P \subseteq \R^{n}_{> - a}\),
    positive parameters \(\varepsilon\),
    \(\eta_{1} \geq \eta_{2} \geq \dots\), and
    \(1/2 \geq \gamma_{1}, \gamma_{2}, \dots\)
    \ENSURE vertices \(x_{t}\) of \(P\) as actions
    \STATE \(\hat{x}_{1} \in P\) arbitrary
    \FOR{\(t=1\) \TO \(T\)}
      \STATE Find distribution \(p_{t}\) with 
        \(\norm[2]{\expectation(x \sim p_{t}){x} - \hat{x}_{t}}
        \leq \gamma_{t} \varepsilon\)
        \label{line:distribution}
        \COMMENT{approximate distribution}
      \STATE \(q_{t} \leftarrow
      (1 - \gamma_{t}) p_{t} + \gamma_{t} \mu\)
      \STATE Sample \(x_{t} \sim q_{t}\).
      \STATE Observe loss
        \(\ell_{t} \coloneqq L_{t}^{\intercal} x_{t}\)
      \STATE
        \(C_{t} \leftarrow
          \expectation(x \sim q_{t}){x x^{\intercal}}
        \)
      \STATE
        \(\hat{L}_{t} \leftarrow
        \ell_{t} C_{t}^{-1} x_{t}\)
      \STATE
        \(y_{t+1, i} \leftarrow
        (a_{i} + \hat{x}_{1, i})^{1 - \eta_{t+1} / \eta_{t}}
        (a_{i} + \hat{x}_{t, i})^{\eta_{t+1} / \eta_{t}}
        \exp (- \eta_{t+1} \hat{L}_{t, i}) - a_{i}\)
         for all \(i \in [n]\)
      \STATE\label{line:projection}
        Find \(\hat{x}_{t+1} \in P\) with
        \(\KLdivergence(a + z, a + \hat{x}_{t+1})
        \leq \KLdivergence(a + z, a + y_{t+1}) + \gamma_{t} \eta_{t+1}\)
        for all \(z \in P\)
        \COMMENT{approximate projection}
    \ENDFOR
  \end{algorithmic}
\end{algorithm}

Now we are ready to state our main theorem
on the regret and complexity of \Algorithm{CombEXP}.

\begin{theorem}[High-probability regret bound for \Algorithm{CombEXP}
  for adaptive adversaries]
  \label{thm:CombEXP}
For \(n \geq 1\) and with the choice
  \begin{align*}
    \gamma_{t} \coloneqq \frac{t^{-1/3}}{2} \qquad \text{ and } \qquad
    \eta_{t} \coloneqq \min \{\gamma_{t}^{2}, \gamma_{t} \lambda\}
  \end{align*}
  Algorithm~\ref{alg:CombEXP} achieves for any time \(T \geq 1\)
  the following regret: With probability at least \(1 - \delta\),
  for any \(x \in P\) we have
  \begin{multline}
    \label{eq:CombEXP}
    \sum_{t=1}^{T} (L_{t}^{\intercal} x_{t} - L_{t}^{\intercal} x)
    \leq
    \left(
      4
      \frac{\KLdivergence(a + x, a + \hat{x}_{1})}{
        \min \{1, 2 \lambda T^{1/3}\}}
      +
      B_{1}
      \sqrt{\frac{3}{\lambda} \ln \frac{n + 2}{\delta}}
      +
      \left(
        (e - 2)
        \frac{\norm[1]{a} + B_{1}}{\lambda}
        +
        2 + \frac{B (B + \varepsilon)}{\lambda}
      \right)
      \frac{3}{4}
    \right) T^{2/3}
    \\
    +
    O \left(
      \max \left\{
        1,
        \frac{B^{2} \max \{(\norm[1]{a} + B_{1}) / \lambda,
          \varepsilon\}}{\lambda},
        \frac{B_{1} \max\{1, B\}}{\lambda},
        \frac{B + \varepsilon}{\sqrt{\lambda}}
      \right\}
      \sqrt{T}
    \right)
    \ln \frac{2 n + 2}{\delta}.
  \end{multline}
  In particular,
  assuming
  \(\alpha - a_{i} \leq z_{i} \leq \beta - a_{i}\)
  for some \(0 < \alpha < \beta\)
  for all \(z \in P\):
\begin{enumerate}
\item \emph{Regret bound}
  We have
  \(\KLdivergence(a + x, a + \hat{x}_{1})
  \leq \frac{4 B^{2}}{\alpha}\)
  so that the upper bound on the regret is proportional to
  \(T^{2/3}\). With probability at least \(1 -
  \delta\), for any \(x \in P\) we have 
\[\sum_{t=1}^{T} (L_{t}^{\intercal} x_{t} - L_{t}^{\intercal} x)  \leq
O\left( 
  \frac{B^2}{\alpha} + \frac{B_1 + (B+\varepsilon)^2 +
      \norm[1]{a}}{\min\{\lambda, \sqrt{\lambda}\}}
    \cdot \ln \frac{2n + 2}{\delta}
  \right) T^{2/3}
 .\]
\item \emph{Complexity.} Using Algorithm~\ref{alg:projection-KL} both
  for projection and distribution generation
  (Lines~\ref{line:distribution} and~\ref{line:projection}) with
  \(\varepsilon = 0\),
  the algorithm makes
  altogether
  \(O \left( \frac{B^{4} \beta}{\alpha^{3} \min \{1, \lambda^{2}\}}
  \right) T^{3}\)
  oracle calls to a linear optimization oracle over \(P\).

  Alternatively using a specialized projection algorithm in
  Line~\ref{line:projection},
  and Algorithm~\ref{alg:linear-decomposition}
  for distribution generation in Line~\ref{line:distribution},
  then Algorithm~\ref{alg:linear-decomposition} calls
  a linear optimization oracle over \(P\)
  at most \(O(B^{2} / \varepsilon^{2}) T^{5/3}\) times
  across all rounds.
\end{enumerate}
\end{theorem}
Obviously, the \(O(B^{2} / \varepsilon^{2}) T^{5/3}\) oracle calls
in the last sentence does not contain the complexity of the
specialized projection algorithm.

Note that the bounds in Theorem~\ref{thm:CombEXP} are \emph{any-time
  guarantees} as the parameters of the algorithm
do not depend on the time horizon \(T\).
The constant factor in the regret bound can be slightly improved
by a more sophisticated choice of the \(\gamma_{t}\) and \(\eta_{t}\),
however, we preferred simple formulae for these parameters.
Just as for \Algorithm{EXP3},
the choice of parameters is different for the best expected regret
and the best high-probability regret.

\subsection{Proof of Theorem~\ref{thm:CombEXP}}
\label{sec:proof-CombEXP}

In this section we will prove Theorem~\ref{thm:CombEXP}. 
We focus on the main regret bound, Equation~\eqref{eq:CombEXP},
the other results easily follow from it.
See Propositions~\ref{prop:projection-KL}
and~\ref{prop:linear-decomposition}
for the complexity of Algorithms~\ref{alg:projection-KL}
and \ref{alg:linear-decomposition}.
The inequality
\(\KLdivergence(a + y, a + \hat{x}_{1}) \leq \frac{4 B^{2}}{\alpha}\)
is derived using \(\ln z \leq z - 1\):
\begin{multline*}
  \KLdivergence(a + y, a + \hat{x}_{1})
  =
  \sum_{i=1}^{n} (a_{i} + y_{i}) \ln \frac{a_{i} + y_{i}}{a_{i} +
    \hat{x}_{1, i}}
  -
  \sum_{i=1}^{n} (a_{i} + y_{i})
  +
  \sum_{i=1}^{n} (a_{i} + \hat{x}_{1, i})
  \\
  \leq
  \sum_{i=1}^{n} (a_{i} + y_{i})
  \left(
    \frac{a_{i} + y_{i}}{a_{i} + \hat{x}_{1, i}}
    -
    1
  \right)
  -
  \sum_{i=1}^{n} (a_{i} + y_{i})
  +
  \sum_{i=1}^{n} (a_{i} + \hat{x}_{1, i})
  =
  \sum_{i=1}^{n}
  \frac{(y_{i} - \hat{x}_{1, i})^{2}}{a_{i} + \hat{x}_{1, i}}
  \leq
  \frac{\norm[2]{y - \hat{x}_{1}}^{2}}{\alpha}
  \leq \frac{4 B^{2}}{\alpha}
  .
\end{multline*}

The proof of Equation~\eqref{eq:CombEXP}
follows the standard approach, whereby
we break-up the regret estimation into various
pieces, which we estimate separately:

\begin{equation}
\label{eq:masterEst}
  \sum_{t=1}^{T} (L_{t}^{\intercal} x_{t} - L_{t}^{\intercal} x)  \leq    
  \underbrace{ \sum_{t=1}^{T}
\left(
    L_{t}^{\intercal} x_{t} - \hat{L}_{t}^{\intercal} \hat{x}_{t}
  \right)}_{\text{Lemma~\ref{lem:estimator-action}}}
+ \underbrace{\sum_{t=1}^{T} (\hat{L}_{t}^{\intercal} \hat{x}_{t}
    - \hat{L}_{t}^{\intercal} x)}_{\text{Lemmas~\ref{lem:EXP}
      and~\ref{lem:EXP-error}}}
  +
\underbrace{
    \sum_{t=1}^{T}
    \left(
      \hat{L}_{t}^{\intercal} x
      - L_{t}^{\intercal} x
    \right)
}_{\text{Lemma~\ref{lem:estimator-baseline-general}}}
.
\end{equation}

Let \(\expectation(t){-} \coloneqq \expectation[x_{1}, L_{1}, \dots,
x_{t-1}, L_{t-1}, L_{t}]{-}\) denote the conditional expectation
operator given the history preceding round \(t\) and also the
adversary's action in round \(t\).
In particular,
\(C_{t} = \expectation(t){x_{t} x_{t}^{\intercal}}
= (1 - \gamma_{t}) P_{t} + \gamma_{t} J\),
with
\(P_{t} \coloneqq \expectation(x \sim p_{t}){x x^{\intercal}}\). 
We first establish some basic bounds on quantities occurring in
Algorithm~\ref{alg:CombEXP}. 

\begin{lemma}[Basic bounds]
  \label{lem:basic-bounds}
  Let \(y \in P\) be arbitrary. 
    \begin{align}
      \label{eq:size-C}
      \norm[2]{C_{t}^{-1}}
      &
      \leq
      \frac{1}{\gamma_{t} \lambda}
      \\
      \label{eq:size-L}
      \norm[2]{\hat{L}_{t}}
      &
      \leq
      \frac{B}{\gamma_{t} \lambda}
      \\
      \label{eq:loss-bound}
      \norm[2]{L_{t}}
      &
      \leq
      \frac{B}{\lambda}
    \end{align}
\begin{proof}
Equation~\eqref{eq:size-C} follows from
\(C_{t} \succeq \gamma_{t} J \succeq \gamma_{t} \lambda I\).
Inequality~\eqref{eq:size-L}
follows via
\begin{equation*}
  \abs{\hat{L}_{t}}
  =
  \abs{\ell_{t} \cdot C_{t}^{-1} x_{t}}
  \leq
  \abs{\ell_{t}} \cdot \norm[2]{C_{t}^{-1}} \cdot \norm[2]{x_{t}}
  \leq
  \frac{B}{\gamma_{t} \lambda}
  .
\end{equation*}
Finally, \eqref{eq:loss-bound} follows from the estimation
\begin{equation*}
  \norm[2]{L_{t}^{\intercal}}
  = \norm[2]{L_{t}^{\intercal} J J^{-1}}
  = \norm[2]*{\expectation(y \sim \mu){L_{t}^{\intercal}
      y y^{\intercal} J^{-1}}}
  \leq
  \expectation(y \sim \mu){\norm[2]{
      L_{t}^{\intercal} y \cdot y^{\intercal} J^{-1}}}
  \leq
  \expectation(y \sim \mu){
    \underbrace{\norm[2]{L_{t}^{\intercal} y}}_{\leq 1}
    \underbrace{\norm[2]{y}}_{\leq B}
    \cdot
    \underbrace{\norm[2]{J^{-1}}}_{\leq 1 / \lambda}}
  \leq
  \frac{B}{\lambda}
  .
  \qedhere
\end{equation*}
\end{proof}
\end{lemma}

We now estimate the pieces of Equation~\eqref{eq:masterEst}. The
following series of upper bounds are independent of the concrete
choice of the parameters \(\gamma_{t}\),
\(\eta_{t}\).
However, for the reader's convenience in the \emph{last inequality} of
each estimation we make the bound explicit by substituting the values
for \(\gamma_{t}\),
\(\eta_{t}\) by the choices given in Theorem~\ref{thm:CombEXP}.
We will tacitly use the following inequality to estimate sums like
\(\sum_{t=1}^{T} \gamma_{t}\):
\begin{equation}
  \label{eq:sum-power}
  \sum_{t=1}^{T} t^{\alpha}
  \leq
  \begin{cases}
    \frac{T^{\alpha + 1} + \alpha}{\alpha + 1}
    \leq
    \frac{T^{\alpha + 1}}{\alpha + 1},
    & -1 < \alpha < 0 \\
    \frac{T^{\alpha + 1} - 1}{\alpha + 1} + T^{\alpha}
    \leq
    \frac{T^{\alpha + 1}}{\alpha + 1} + T^{\alpha}
    , & \alpha > 0
  \end{cases}
\end{equation}

We first estimate the regret when using the loss estimators
\(\hat L_t\).
For this we use a generalized variant of EXP (see
Lemma~\ref{lem:EXP3}), which works with
arbitrary convex sets contained in the positive orthant. 

\begin{lemma}
  \label{lem:EXP}
  \begin{equation}
    \label{eq:EXP-simplified}
   \begin{split}
    \sum_{t=1}^{T} (\hat{L}_{t}^{\intercal} \hat{x}_{t}
    - \hat{L}_{t}^{\intercal} x)
    &
    \leq
    \frac{\KLdivergence(a + x, a + \hat{x}_{1})}{\eta_{T}}
    +
    \sum_{t=1}^{T-1} \gamma_{t}
    +
    (e - 2)
    \sum_{t=1}^{T} \eta_{t}
    \sum_{i=1}^{n} (a_{i} + \hat{x}_{t, i}) \hat{L}_{t, i}^{2}
    \\
    &
    \leq
    4
    \frac{\KLdivergence(a + x, a + \hat{x}_{1}) T^{2/3}}{
      \min \{1, 2 \lambda T^{1/3}\}}
    +
    \frac{3}{4} T^{2/3}
    +
    (e - 2)
    \sum_{t=1}^{T} \eta_{t}
    \sum_{i=1}^{n} (a_{i} + \hat{x}_{t, i}) \hat{L}_{t, i}^{2}
    .
   \end{split}
  \end{equation}
\begin{proof}
This follows from Lemma~\ref{lem:EXP3} with the \(\hat{L}_{t}\) as
loss vectors and the \(a + \hat{x}_{t}\) as played actions.
Note that \(a\) cancels on the left hand side in
\((\hat{L}_{t}^{\intercal} (a + \hat{x}_{t})
- \hat{L}_{t}^{\intercal} (a + x))\).
\end{proof}
\end{lemma}

In a next step we estimate the last term of
Equation~\eqref{eq:EXP-simplified}.

\begin{lemma}
\label{lem:EXP-error}
  With probability at least \(1-\delta\)
\begin{equation}
  \label{eq:EXP-error}
 \begin{split}
  \sum_{t=1}^{T}
  \eta_{t}
  \sum_{i=1}^{n} (a_{i} + \hat{x}_{t, i}) \hat{L}_{t, i}^{2}
  &
  \leq
  \frac{\norm[1]{a} + B_{1}}{\lambda}
  \sum_{t=1}^{T} \frac{\eta_{t}}{\gamma_{t}}
  + \frac{(\norm[1]{a} + B_{1}) B^{2}}{\lambda^{2}} \sqrt{\frac{1}{2}
    \sum_{t=1}^{T}
    \frac{\eta_{t}^{2}}{\gamma_{t}^{4}} \cdot \ln \frac{1}{\delta}}
  \\
  &
  \leq
  \frac{\norm[1]{a} + B_{1}}{\lambda}
  \frac{3}{4} T^{2/3}
  +
  \frac{(\norm[1]{a} + B_{1}) B^{2}}{\lambda^{2}}
  \sqrt{\frac{1}{2} T \ln \frac{1}{\delta}}
  .
 \end{split}
\end{equation}
\begin{proof}
This is a special case of the Azuma–Hoeffding inequality
(recalled in Theorem~\ref{thm:Azuma})
using the bounds
\begin{equation*}
  0
  \leq
  \sum_{i=1}^{n} (a_{i} + \hat{x}_{t, i}) \hat{L}_{t, i}^{2}
  \leq
  \sum_{i=1}^{n} (a_{i} + \hat{x}_{t, i}) \left(
    \frac{B}{\gamma_{t} \lambda}
  \right)^{2}
  \leq
  \frac{(\norm[1]{a} + B_{1}) B^{2}}{(\gamma_{t} \lambda)^{2}}
\end{equation*}
and
\begin{multline*}
  \expectation(t){\sum_{i=1}^{n}
    (a_{i} + \hat{x}_{t, i}) \hat{L}_{t, i}^{2}}
  =
  \expectation(t){\sum_{i=1}^{n} (a_{i} + \hat{x}_{t, i})
    \ell_{t}^{2}
    e_{i}^{\intercal} C_{t}^{-1} x_{t}
    x_{t}^{\intercal} C_{t}^{-1} e_{i}}
  =
  \sum_{i=1}^{n} (a_{i} + \hat{x}_{t, i})
  \ell_{t}^{2}
  e_{i}^{\intercal} C_{t}^{-1} C_{t} C_{t}^{-1} e_{i}
  \\
  \leq
  \sum_{i=1}^{n} (a_{i} + \hat{x}_{t, i})
  e_{i}^{\intercal} C_{t}^{-1} e_{i}
  \leq
  \frac{\norm[1]{a} + B_{1}}{\gamma_{t} \lambda}
  .
  \qedhere
\end{multline*}
\end{proof}
\end{lemma}

Next we bound the difference between the true loss
\(L_t^\intercal x_t\)
and the expected estimated loss \(\hat L_t^\intercal \hat x_t\).

\begin{lemma}
\label{lem:estimator-action}
  With probability at least \(1-\delta\)
\begin{multline}
  \label{eq:estimator-action}
  \sum_{t=1}^{T}
  \left(
    L_{t}^{\intercal} x_{t} - \hat{L}_{t}^{\intercal} \hat{x}_{t}
  \right)
  \leq
  \left(
    1 + \frac{B (B + \varepsilon)}{\lambda}
  \right)
  \sum_{t=1}^T \gamma_{t}
  +
  \sqrt{2 \left(
      T
      +
      \frac{(3 B + \varepsilon) (B + \varepsilon)}{\lambda}
      \sum_{t=1}^{T} \frac{\gamma_{t}}{(1 - \gamma_{t})^{2}}
    \right)
    \ln \frac{1}{\delta}}
  \\
  +
  \frac{1}{3}
  \left(
    \frac{B}{\sqrt{\gamma_{T} (1 - \gamma_{T}) \lambda}}
    + 3 + \frac{B^{2} \varepsilon}{\lambda}
  \right)
  \ln \frac{1}{\delta}
  \\
  \leq
  \left(
    1 + \frac{B (B + \varepsilon)}{\lambda}
  \right)
  \sum_{t=1}^T \gamma_{t}
  +
  \frac{B}{3 \sqrt{\gamma_{T} (1 - \gamma_{T}) \lambda}}
  \ln \frac{1}{\delta}
  +
  \left[
    3 + \frac{B^{2} \varepsilon}{\lambda}
    +
    O \left(
      \max \left\{
        1, \frac{B + \varepsilon}{\sqrt{\lambda}}
      \right\}
      \sqrt{T}
    \right)
  \right]
  \max \left\{ 1, \ln \frac{1}{\delta} \right\}
  \\
  \leq
  \left(
    1 + \frac{B (B + \varepsilon)}{\lambda}
  \right)
  \frac{3}{4} T^{2/3}
  +
  \frac{\sqrt{2} B}{3 \sqrt{\lambda}}
  (T^{1/6} + T^{-1/6})
  \ln \frac{1}{\delta}
  +
  \left[
    3 + \frac{B^{2} \varepsilon}{\lambda}
    +
    O \left(
      \max \left\{
        1, \frac{B + \varepsilon}{\sqrt{\lambda}}
      \right\}
      \sqrt{T}
    \right)
  \right]
  \max \left\{ 1, \ln \frac{1}{\delta} \right\}
  .
\end{multline}
\begin{proof}
Let \(\tilde{x}_{t} \coloneqq \expectation(x \sim p_{t}){x}\)
and
\(\overline{x_{t}} \coloneqq \expectation(t){x_{t}}
= (1 - \gamma_{t}) \tilde{x}_{t} + \gamma_{t} u\).
As also
\(\norm[2]{\hat{x}_{t} - \tilde{x}_{t}} \leq \gamma_{t} \varepsilon\),
we have
\(\overline{x_{t}} = (1 - \gamma_{t}) \hat{x}_{t} + \gamma_{t} v\)
for \(v \coloneqq u + \frac{1 - \gamma_{t}}{\gamma_{t}}
(\tilde{x}_{t} - \hat{x}_{t})\)
with
\(\norm[2]{v} \leq B + \varepsilon (1 - \gamma_{t})
\leq B + \varepsilon\).
We consider the martingale difference sequence
\begin{equation*}
  X_{t} \coloneqq
  L_{t}^{\intercal} x_{t} - \hat{L}_{t}^{\intercal} \hat{x}_{t}
  - \expectation(t){L_{t}^{\intercal} x_{t}
    - \hat{L}_{t}^{\intercal} \hat{x}_{t}}
  =
  L_{t}^{\intercal} x_{t} - \hat{L}_{t}^{\intercal} \hat{x}_{t}
  - L_{t}^{\intercal} \overline{x_{t}}
  + L_{t}^{\intercal} \hat{x}_{t}
  =
  L_{t}^{\intercal} x_{t} - \hat{L}_{t}^{\intercal} \hat{x}_{t}
  + \gamma_{t} L_{t}^{\intercal} (\hat{x}_{t} - v).
\end{equation*}
Note that
as
\(\tilde{x}_{t} \tilde{x}_{t}^{\intercal}
= \expectation(x \sim p_{t}){x}
\expectation(x \sim p_{t}){x}^{\intercal}
\preceq \expectation(x \sim p_{t}){x x^{\intercal}} = P_{t}
\leq C_{t} / (1 - \gamma_{t})\)
\begin{equation*}
  \left(
    \hat{L}_{t}^{\intercal} \tilde{x}_{t}
  \right)^{2}
  =
  \hat{L}_{t}^{\intercal} \tilde{x}_{t}
  \tilde{x}_{t}^{\intercal} \hat{L}_{t}
  =
  \ell_{t}^{2}
  x_{t}^{\intercal} C_{t}^{-1} \tilde{x}_{t}
  \tilde{x}_{t}^{\intercal} C_{t}^{-1} x_{t}
  \leq
  x_{t}^{\intercal} C_{t}^{-1} P_{t} C_{t}^{-1} x_{t}
  \leq
  \frac{x_{t}^{\intercal} C_{t}^{-1} x_{t}}{1 - \gamma_{t}}
  \leq
  \frac{B^{2}}{\gamma_{t} \lambda (1 - \gamma_{t})}
  \leq
  \frac{B^{2}}{\gamma_{T} \lambda (1 - \gamma_{T})}
  ,
\end{equation*}
and
\begin{equation*}
  \abs{
    \hat{L}_{t}^{\intercal} \hat{x}_{t}
    -
    \hat{L}_{t}^{\intercal} \tilde{x}_{t}}
  \leq
  \norm[2]{\hat{L}_{t}} \cdot
  \norm[2]{\hat{x}_{t} - \tilde{x}_{t}}
  \leq
  \frac{B^{2}}{\gamma_{t} \lambda}
  \gamma_{t} \varepsilon
  =
  \frac{B^{2} \varepsilon}{\lambda}
  ,
\end{equation*}
hence
\begin{equation*}
  \abs{X_{t}}
  \leq
  \abs*{L_{t}^{\intercal} x_{t}}
  + \abs*{L_{t}^{\intercal} \overline{x_{t}}}
  + \abs*{L_{t}^{\intercal} \hat{x}_{t}}
  + \abs*{\hat{L}_{t}^{\intercal} \hat{x}_{t}
    - \hat{L}_{t}^{\intercal} \tilde{x}_{t}}
  + \abs*{\hat{L}_{t}^{\intercal} \tilde{x}_{t}}
  \leq
  3
  + \frac{B^{2} \varepsilon}{\lambda}
  + \frac{B}{\sqrt{\gamma_{T} (1 - \gamma_{T}) \lambda}}
  ,
\end{equation*}
and the variance of \(X_{t}\) is easily bounded by:
\begin{multline*}
  \variance(t){X_{t}}
  \leq
  \expectation(t){(L_{t}^{\intercal} x_{t}
    - \hat{L}_{t}^{\intercal} \hat{x}_{t})^{2}}
  =
  \expectation(t){(\ell_{t} \cdot
    (1 - x_{t}^{\intercal} C_{t}^{-1} \hat{x}_{t}))^{2}}
  \\
  \leq
  \expectation(t){(1 - x_{t}^{\intercal} C_{t}^{-1} \hat{x}_{t})^{2}}
  =
  \expectation(t){1 - 2 x_{t}^{\intercal} C_{t}^{-1} \hat{x}_{t}
    +
    \hat{x}_{t}^{\intercal} C_{t}^{-1} x_{t} x_{t}^{\intercal}
    C_{t}^{-1} \hat{x}_{t}}
  \\
  =
  1 - 2 \overline{x_{t}}^{\intercal} C_{t}^{-1} \hat{x}_{t}
  +
  \hat{x}_{t}^{\intercal} C_{t}^{-1} \hat{x}_{t}
  =
  1
  -
  \frac{1 - 2 \gamma_{t}}{(1 - \gamma_{t})^{2}}
  \overline{x_{t}}^{\intercal} C_{t}^{-1} \overline{x_{t}}
  + \frac{\gamma_{t}^{2}}{(1 - \gamma_{t})^{2}} \left(
    v - 2 \overline{x_{t}}
  \right)^{\intercal} C_{t}^{-1} v
  \leq
  1 + \frac{\gamma_{t} (3B + \varepsilon)(B + \varepsilon)}{
    (1 - \gamma_{t})^{2} \lambda}
  .
\end{multline*}
Hence
Benett's inequality (Theorem~\ref{thm:Benett},
\cite[(18)]{HoeffdingSuper2012})
applied to the martingale difference sequence \(X_{t}\)
provides
\begin{equation*}
  \sum_{t=1}^{T}
  \left(
    L_{t}^{\intercal} x_{t} - \hat{L}_{t}^{\intercal} \hat{x}_{t}
    + \gamma_{t} L_{t}^{\intercal} (\hat{x}_{t} - v)
  \right)
  \leq
  \begin{aligned}[t]
  & \frac{1}{3}
  \left(
    \frac{B}{\sqrt{\gamma_{T} (1 - \gamma_{T}) \lambda}}
    + 3 + \frac{B^{2} \varepsilon}{\lambda}
  \right)
  \ln \frac{1}{\delta}
  \\
  &
  +
  \sqrt{2 \left(
      T
      +
      \frac{(3 B + \varepsilon) (B + \varepsilon)}{\lambda}
      \sum_{t=1}^{T} \frac{\gamma_{t}}{(1 - \gamma_{t})^{2}}
    \right)
    \ln \frac{1}{\delta}}
  .  
  \end{aligned}
\end{equation*}
The claim follows by using
\(\abs{L_{t}^{\intercal} (\hat{x}_{t} - v)}
\leq 1 + B (B + \varepsilon) / \lambda\).
\end{proof}
\end{lemma}

Finally, we bound the difference between the true loss \(
L_{t}^{\intercal} x\) and the estimated loss \(\hat{L}_{t}^{\intercal}
x\) for any point \(x \in P\). 

\begin{lemma}
\label{lem:estimator-baseline-general}
  For all \(0 < \delta < 1\)
  with probability at least \(1-\delta\)
  for every \(x \in \R^{n}\) simultaneously
  \begin{equation}
    \label{eq:estimator-baseline-general}
    \sum_{t=1}^{T}
    \left(
      \hat{L}_{t}^{\intercal} x
      - L_{t}^{\intercal} x
    \right)
    \leq
    \frac{\norm[1]{x}}{3}
    \left(
      \frac{B}{\lambda} + \frac{1}{\gamma_{T} \lambda}
    \right)
    \ln \frac{2 n}{\delta}
    +
    \norm[1]{x}
    \sqrt{\frac{2}{\lambda}
      \sum_{t=1}^{T} \frac{1}{\gamma_{t}}
      \ln \frac{2 n}{\delta}}
    .
  \end{equation}
  In particular, with probability at least \(1 - \delta\),
  for all \(x \in P\) simultaneously
  \begin{equation}
    \label{eq:estimator-baseline}
    \sum_{t=1}^{T}
    \left(
      \hat{L}_{t}^{\intercal} x
      - L_{t}^{\intercal} x
    \right)
    \leq
    \frac{B_{1}}{3 \lambda}
    \left(
      B + 2 T^{1/3}
    \right)
    \ln \frac{2 n}{\delta}
    +
    B_{1}
    T^{2/3} \sqrt{1 + \frac{4}{3 T}}
    \sqrt{\frac{3}{\lambda} \ln \frac{2 n}{\delta}}
    .
  \end{equation}
  \begin{remark}
    Restricting the statement for all \(x \geq 0\),
    the \(\ln (2 n / \delta)\)
    can be replaced by \(\ln (n / \delta)\).
  \end{remark}
\begin{proof}
Let \(x = \pm e_{i}\) be a coordinate vector or its negation.
Then
\begin{equation*}
  \variance(t){\hat{L}_{t}^{\intercal} x - L_{t}^{\intercal} x}
  \leq
  \expectation(t){(\hat{L}_{t}^{\intercal} x)^{2}}
  \leq
  \expectation(t){x^{\intercal} C_{t}^{-1} x_{t} x_{t}^{\intercal}
    C_{t}^{-1} x}
  =
  x^{\intercal} C_{t}^{-1} x
  \leq
  \frac{1}{\gamma_{t} \lambda}
  ,
\end{equation*}
and
\begin{equation*}
  \abs{\hat{L}_{t}^{\intercal} x - L_{t}^{\intercal} x}
  \leq
  \frac{B}{\lambda} + \frac{1}{\gamma_{t} \lambda}
  .
\end{equation*}
Hence by Benett's inequality
(Theorem~\ref{thm:Benett}, \cite[(18)]{HoeffdingSuper2012})
the claim follows
for a fixed vector \(x = \pm e_{i}\)
with probability at least \(1 - \delta / (2 n)\).
Hence by the union bound,
it holds for all \(x = \pm e_{i}\) simultaneously
with probability at least \(1 - \delta\).
Finally, the inequality for a general \(x\)
follows by taking linear combinations with the
absolute values of the coefficients of
\(x\).
\end{proof}
\end{lemma}

Summing up \eqref{eq:EXP-simplified},
\eqref{eq:EXP-error},
\eqref{eq:estimator-action}
(substituting \(\delta / (2 n + 2)\) for \(\delta\) in the latter two)
 and \eqref{eq:estimator-baseline}
(substituting \(2 n \delta / (2 n + 2)\) for \(\delta\)),
with probability at least \(1 - \delta\)
yields \eqref{eq:CombEXP} of
Theorem~\ref{thm:CombEXP}.

\section{A high-probability regret bound for
  \Algorithm{ComBand}}
\label{sec:regret-bound-ComBandit}

In this section we will show that \Algorithm{ComBand} of
\cite{ComBandit2012} achieves a high-probability regret bound of
\(O(T^{2/3})\)
without any modifications. While this is worse than the optimal regret
of \(O(\sqrt{T})\)
obtained by \Algorithm{GeometricHedge} in
\cite{RegretLinearBandit2008}, it shows that already
Algorithm~\ref{alg:ComBandit}, the vanilla
version of \Algorithm{ComBand} without any correction terms
suffices to achieve a high-probability regret bound.

\begin{algorithm}
  \caption{\Algorithm{ComBand}}
  \label{alg:ComBandit}
  \begin{algorithmic}[1]
    \REQUIRE Losses \(L_{t}\), action set \(A \subseteq \R^{n}\),
     positive parameters \(\eta_{1} \geq \eta_{2} \geq \dots\),
     \(1/2 \geq \gamma_{1} \geq \gamma_{2}\geq \dots\)
    \ENSURE actions \(x_{t} \in A\)
    \FOR{\(t=1\) \TO \(T\)}
      \STATE \(w_{t}(x) \leftarrow \sum_{i=1}^{t-1}
        \hat{L}_{i}^{\intercal} x\) \quad for all \(x\)
      \STATE \(W_{t} \leftarrow \sum_{x} w_{t}(x)\)
      \STATE \(p_{t}(x) \leftarrow w_{t}(x) / W_{t}\)
        \quad for all \(x\)
      \STATE
        \(q_{t} \leftarrow (1 - \gamma_{t}) p_{t} + \gamma_{t} \mu\)
      \STATE Sample \(x_{t} \sim q_{t}\).
      \STATE Observe loss
        \(\ell_{t} \coloneqq L_{t}^{\intercal} x_{t}\).
      \STATE
        \(C_{t} \leftarrow
          \expectation(x \sim q_{t}){x x^{\intercal}}
        \)
      \STATE
        \(\hat{L}_{t} \leftarrow
        \ell_{t} C_{t}^{-1} x_{t}\)
    \ENDFOR
  \end{algorithmic}
\end{algorithm}

\begin{theorem}
  \label{thm:ComBandit}
  With the choice
  \begin{align*}
    \eta_{t} \coloneqq \frac{\gamma_{t} \lambda}{B^{2}}
    \qquad \text{ and } \qquad 
    \gamma_{t} \coloneqq
    \frac{t^{-1/3}}{2}
  \end{align*}
  Algorithm~\ref{alg:ComBandit} achieves regret
  \begin{align}
    \label{eq:ComBandit}
    \left(
      \frac{\sqrt{3} B}{\sqrt{\lambda}}
      \sqrt{\ln \frac{N + 2}{\delta}}
      +
      n \frac{3 (e - 2) \lambda}{4 B^{2}}
      +
      \frac{3}{2}
    \right)
    T^{2/3}
    & +
    O \left(
      n \frac{\lambda}{B^{2}}
      +
      \left(
        1 + \frac{B^{2}}{\lambda}
      \right)
      \ln \frac{N + 2}{\delta}
    \right)
    \sqrt{T} \\
    & \leq 
    O \left(
      \frac{B}{\sqrt{\lambda}}
      \sqrt{\ln \frac{N + 2}{\delta}}
      +
      n \frac{\lambda}{B^{2}}
    \right)
    T^{2/3}
  \end{align}
  with probability at least \(1 - \delta\)
  for \(0 < \delta < 1\).
\end{theorem}
\begin{remark}
  Similar to Theorem~\ref{thm:CombEXP},
  it is possible to change the \(\ln ((N+2) / \delta)\)
  in the coefficient of \(T^{2/3}\)
  to the possibly much smaller \(\ln ((n+2) / \delta)\)
  with a suitable altering of the other constants.
  However,
  since an \(T^{1/3} \ln N\) term will still remain
  in the regret bound,
  this does not seem to be a significant improvement.
\end{remark}

We use the same notation as in Section~\ref{sec:proof-CombEXP}
for \Algorithm{CombEXP},
which we recall here for the reader's convenience.
Let \(\expectation(t){-} \coloneqq \expectation[x_{1}, L_{1}, \dots,
x_{t-1}, L_{t-1}, L_{t}]{-}\) denote the conditional expectation
operator given the history preceding round \(t\) and also the
adversary's action in round \(t\).
Let
\(\tilde{x}_{t} \coloneqq \expectation(x \sim p_{t}){x}\)
and
\(P_{t} \coloneqq
\expectation(x \sim p_{t}){x x^{\intercal}}
\)
denote the expectation and variance of distribution \(p_{t}\),
respectively.
Note that
\(C_{t}
=
\expectation(t){x_{t} x_{t}^{\intercal}}
=
(1 - \gamma_{t}) P_{t} + \gamma_{t} J\).

\begin{lemma}[Basic bounds]
  \label{lem:basic-bounds-Comb}
  Let \(x\), \(y_{1}\), and \(y_{2}\) be arbitrary actions.
  \begin{enumerate}
  \item Bounds on size
    \begin{align}
      \label{eq:size-C-Comb}
      \abs{y_{1}^{\intercal} C_{t}^{-1} y_{2}}
      &
      \leq
      \frac{B^{2}}{\gamma_{t} \lambda}
      \\
      \label{eq:size-L-Comb}
      \abs{\hat{L}_{t}^{\intercal} x}
      &
      \leq
      \frac{B^{2}}{\gamma_{t} \lambda}
    \end{align}
  \item Bounds on expectation
    \begin{align}
      \label{eq:expect-C}
      \expectation(t){x_{t}^{\intercal} C_{t}^{-1} x_{t}}
      &
      = n
      \\
    \end{align}
  \end{enumerate}
\begin{proof}
Equation~\eqref{eq:size-C-Comb} follows from the bounds
\(\norm[2]{y_{1}}, \norm[2]{y_{2}} \leq B\)
and \(\norm[2]{C_{t}^{-1}} \leq 1 / (\gamma_{t} \lambda)\),
as \(C_{t} \succeq \gamma_{t} J \succeq \gamma_{t} \lambda I\).
Inequality~\eqref{eq:size-L-Comb}
follows via
\begin{equation*}
  \abs{\hat{L}_{t}^{\intercal} x}
  =
  \abs{\ell_{t} \cdot x_{t}^{\intercal} C_{t}^{-1} x}
  \leq
  \frac{B^{2}}{\gamma_{t} \lambda}
  .
\end{equation*}

To prove \eqref{eq:expect-C},
we use a trick using the trace function to compute the expectation:
\begin{equation*}
  \expectation(t){x_{t}^{\intercal} C_{t}^{-1} x_{t}}
  =
  \expectation(t){\tr(C_{t}^{-1} x_{t} x_{t}^{\intercal})}
  = \tr(C_{t}^{-1} C_{t})
  = \tr(I) = n
  .
\end{equation*}
\end{proof}
\begin{remark}
  One can similarly prove
  \(\expectation(y \sim p_{t}){y^{\intercal} P_{t}^{-1} y} = n\),
  but it will not be used in the following.
\end{remark}
\end{lemma}

As in the case of \Algorithm{CombEXP},
the lemmas below are independent of the choice of
the \(\gamma_{t}\), \(\eta_{t}\)
except for the last formula in each lemma,
where we particularize the bounds by substituting parameters.

First instead of the real regret,
we estimate the regret computed using the estimators
\(\hat{L}_{t}\).
\begin{lemma}
  \label{lem:EXP-Comb}
  With probability at least \(1 - \delta\)
  \begin{equation}
   \begin{split}
    \label{eq:EXP-Comb}
    \sum_{t=1}^{T} (\hat{L}_{t}^{\intercal} \tilde{x}_{t}
    - \hat{L}_{t}^{\intercal} x)
    &
    \leq
    \frac{\ln N}{\eta_{T}}
    + (e - 2) \left(
      n \sum_{t=1}^{T} \frac{\eta_{t}}{1 - \gamma_{t}}
      + \frac{B^{2}}{\lambda} \sqrt{\frac{1}{2}
        \sum_{t=1}^{T}
        \frac{\eta_{t}^{2}}{\gamma_{t}^{2} (1 - \gamma_{t})^{2}}
        \cdot \ln \frac{1}{\delta}}
    \right)
    \\
    &
    \leq
    \frac{2 B^{2} \ln N}{\lambda} T^{1/3}
    + (e - 2) \left(
      n \frac{3 \lambda}{4 B^{2}} (T^{2/3} + 2 T^{1/3})
      +
      \frac{B}{\sqrt{\lambda}} \sqrt{2 T \ln \frac{1}{\delta}}
    \right)
    .
   \end{split}
  \end{equation}
\begin{proof}
By Lemma~\ref{lem:EXP3},
\begin{equation}
  \label{eq:EXP}
  \sum_{t=1}^{T}
  \left(
    \hat{L}_{t}^{\intercal} \tilde{x}_{t} - \hat{L}_{t}^{\intercal} x
  \right)
  \leq
  \frac{\ln N}{\eta_{T}}
  + (e - 2) \sum_{t=1}^{T} \eta_{t}
  \expectation(y \sim p_{t}){(\hat{L}_{t}^{\intercal} y)^{2}}
  .
\end{equation}
To estimate the last term,
first note that
\begin{equation*}
  \expectation(y \sim p_{t}){(\hat{L}_{t}^{\intercal} y)^{2}}
  =
  \expectation(y \sim p_{t}){\hat{L}_{t}^{\intercal}
    y y^{\intercal} \hat{L}_{t}}
  =
  \hat{L}_{t}^{\intercal} P_{t} \hat{L}_{t}
  =
  \ell_{t}^{2} x_{t}^{T} C_{t}^{-1} P_{t} C_{t}^{-1} x_{t}
  \leq
  \frac{x_{t}^{\intercal} C_{t}^{-1} x_{t}}{1 - \gamma_{t}}
  .
\end{equation*}
So far combining our estimates provides
\begin{equation}
  \label{eq:EXP-simplified-Comb}
  \sum_{t=1}^{T} (\hat{L}_{t}^{\intercal} \tilde{x}_{t}
  - \hat{L}_{t}^{\intercal} x)
  \leq
  \frac{\ln N}{\eta_{T}}
  + \sum_{t=1}^{T} \eta_{t}
  \frac{x_{t}^{\intercal} C_{t}^{-1} x_{t}}{1 - \gamma_{t}}
  =
  \frac{2 B^{2} \ln N}{\lambda} T^{1/3}
  + (e - 2) \sum_{t=1}^{T} \eta_{t}
  \frac{x_{t}^{\intercal} C_{t}^{-1} x_{t}}{1 - \gamma_{t}}
  .
\end{equation}
To estimate the last term on the right-hand side,
we apply the Azuma–Hoeffding inequality
using \eqref{eq:size-C-Comb} and \eqref{eq:expect-C}
for bounding the summands and their expectation,
which readily proves the lemma:
\begin{equation*}
  \sum_{t=1}^{T} \eta_{t}
  \frac{x_{t}^{\intercal} C_{t}^{-1} x_{t}}{1 - \gamma_{t}}
  \leq
  n \sum_{t=1}^{T} \frac{\eta_{t}}{1 - \gamma_{t}}
  + \frac{B^{2}}{\lambda} \sqrt{\frac{1}{2}
    \sum_{t=1}^{T}
    \frac{\eta_{t}^{2}}{\gamma_{t}^{2} (1 - \gamma_{t})^{2}}
    \cdot \ln \frac{1}{\delta}}
  .
  \qedhere
\end{equation*}
\end{proof}
\end{lemma}

We turn our attention to the difference between the real loss vectors
\(L_{t}\) and their estimators \(\hat{L}_{t}\).
We start by comparing the loss of the played action.
\begin{lemma}
  With probability at least \(1-\delta\)
\begin{equation}
 \begin{split}
  \label{eq:estimator-action-Comb}
  \sum_{t=1}^{T}
  \left(
    L_{t}^{\intercal} x_{t} - \hat{L}_{t}^{\intercal} \tilde{x}_{t}
  \right)
  &
  \leq
  2 \sum_{t=1}^{T} \gamma_{t}
  +
  \frac{1}{3}
  \left(
    2 + \frac{B}{\sqrt{\gamma_{T} \lambda (1 - \gamma_{T})}}
  \right)
  \ln \frac{1}{\delta}
  +
  \sqrt{2 \left(
      T
      +
      \frac{3  B^{2}}{ \lambda}
      \sum_{t=1}^{T} \frac{\gamma_{t}}{(1 - \gamma_{t})^{2}}
    \right)
    \ln \frac{1}{\delta}}
  \\
  &
  \leq
  \frac{3}{2} T^{2/3}
  +
  \frac{1}{3}
  \left(
    2
    +
    \frac{\sqrt{2} B}{\sqrt{\lambda}}
    (T^{1/6} + T^{-1/6})
  \right)
  \ln \frac{1}{\delta}
  +
  \sqrt{2 \left(
      T + \frac{9 B^{2}}{\lambda} T^{2/3}
    \right)
    \ln \frac{1}{\delta}}
  .
 \end{split}
\end{equation}
\begin{proof}
Let \(\overline{x_{t}} \coloneqq \expectation(t){x_{t}}
= (1 - \gamma_{t}) \tilde{x}_{t} + \gamma_{t} u\)
denote the conditional expectation of \(x_{t}\)
given the history before round \(t\) and loss \(L_{t}\).
The statement is a special case of
Benett's inequality (see Theorem~\ref{thm:Benett})
for the martingale
\begin{equation*}
  X_{t} \coloneqq
  L_{t}^{\intercal} x_{t} - \hat{L}_{t}^{\intercal} \tilde{x}_{t}
  - \expectation(t){L_{t}^{\intercal} x_{t}
    - \hat{L}_{t}^{\intercal} \tilde{x}_{t}}
  =
  L_{t}^{\intercal} x_{t} - \hat{L}_{t}^{\intercal} \tilde{x}_{t}
  - L_{t}^{\intercal} \overline{x_{t}}
  + L_{t}^{\intercal} \tilde{x}_{t}
  =
  L_{t}^{\intercal} x_{t} - \hat{L}_{t}^{\intercal} \tilde{x}_{t}
  + \gamma_{t} L_{t}^{\intercal} (\tilde{x}_{t} - u).
\end{equation*}
Note that
\(\tilde{x}_{t} \tilde{x_{t}}^{\intercal} \preceq P_{t}\)
by Jensen's inequality,
therefore
\begin{equation*}
  \left(
    \hat{L}_{t}^{\intercal} \tilde{x}_{t}
  \right)^{2}
  =
  \hat{L}_{t}^{\intercal} \tilde{x}_{t}
  \tilde{x}_{t}^{\intercal} \hat{L}_{t}
  =
  \ell_{t}^{2}
  x_{t}^{\intercal} C_{t}^{-1} \tilde{x}_{t}
  \tilde{x}_{t}^{\intercal} C_{t}^{-1} x_{t}
  \leq
  x_{t}^{\intercal} C_{t}^{-1} P_{t} C_{t}^{-1} x_{t}
  \leq
  \frac{x_{t}^{\intercal} C_{t}^{-1} x_{t}}{1 - \gamma_{t}}
  \leq
  \frac{B^{2}}{\gamma_{t} \lambda (1 - \gamma_{t})}
  ,
\end{equation*}
hence
\begin{equation*}
  \abs{X_{t}}
  \leq
  1 + \frac{B}{\sqrt{\gamma_{t} \lambda (1 - \gamma_{t})}}
  + 2 \gamma_{t}
  \leq
  2 + \frac{B}{\sqrt{\gamma_{T} \lambda (1 - \gamma_{T})}}
  ,
\end{equation*}
and the variance of \(X_{t}\) is easily bounded by:
\begin{multline*}
  \variance(t){X_{t}}
  \leq
  \expectation(t){(L_{t}^{\intercal} x_{t}
    - \hat{L}_{t}^{\intercal} \tilde{x}_{t})^{2}}
  =
  \expectation(t){(\ell_{t} \cdot
    (1 - x_{t}^{\intercal} C_{t}^{-1} \tilde{x}_{t}))^{2}}
  \\
  \leq
  \expectation(t){(1 - x_{t}^{\intercal} C_{t}^{-1} \tilde{x}_{t})^{2}}
  =
  \expectation(t){1 - 2 x_{t}^{\intercal} C_{t}^{-1} \tilde{x}_{t}
    +
    \tilde{x}_{t}^{\intercal} C_{t}^{-1} x_{t} x_{t}^{\intercal}
    C_{t}^{-1} \tilde{x}_{t}}
  \\
  =
  1 - 2 \overline{x_{t}}^{\intercal} C_{t}^{-1} \tilde{x}_{t}
  +
  \tilde{x}_{t}^{\intercal} C_{t}^{-1} \tilde{x}_{t}
  =
  1
  -
  \frac{1 - 2 \gamma_{t}}{(1 - \gamma_{t})^{2}}
  \overline{x_{t}}^{\intercal} C_{t}^{-1} \overline{x_{t}}
  + \frac{\gamma_{t}^{2}}{(1 - \gamma_{t})^{2}} \left(
    u - 2 \overline{x_{t}}
  \right)^{\intercal} C_{t}^{-1} u
  \leq
  1 + \frac{3 \gamma_{t} B^{2}}{(1 - \gamma_{t})^{2} \lambda}
  .
\end{multline*}
Benett's inequality provides
\begin{equation*}
  \sum_{t=1}^{T}
  \left(
    L_{t}^{\intercal} x_{t} - \hat{L}_{t}^{\intercal} \tilde{x}_{t}
    + \gamma_{t} L_{t}^{\intercal} (x_{t} - u)
  \right)
  \leq
  \frac{1}{3}
  \left(
    2 + \frac{B}{\sqrt{\gamma_{T} \lambda (1 - \gamma_{T})}}
  \right)
  \ln \frac{1}{\delta}
  +
  \sqrt{2 \left(
      T
      +
      \frac{3 B^{2}}{\lambda}
      \sum_{t=1}^{T} \frac{\gamma_{t}}{(1 - \gamma_{t})^{2}}
    \right)
    \ln \frac{1}{\delta}}
  .
\end{equation*}
The claim follows by using
\(\abs{L_{t}^{\intercal} (x_{t} - u)} \leq 2\).
\end{proof}
\end{lemma}

Now we compare the losses \(L_{t}\)
with their estimator \(\hat{L}_{t}\)
for all fixed actions.
\begin{lemma}
  For all \(0 < \delta < 1\)
  with probability at least \(1-\delta\)
  for every \(x \in A\) simultaneously
  \begin{equation}
   \begin{split}
    \label{eq:estimator-baseline-Comb}
    \sum_{t=1}^{T}
    \left(
      \hat{L}_{t}^{\intercal} x
      - L_{t}^{\intercal} x
    \right)
    &
    \leq
    \frac{1}{3}
    \left(
      1 + \frac{B^{2}}{\gamma_{T} \lambda}
    \right)
    \ln \frac{N}{\delta}
    +
    \sqrt{\frac{2 B^{2}}{\lambda}
      \sum_{t=1}^{T} \frac{1}{\gamma_{t}}
      \cdot \ln \frac{N}{\delta}}
    \\
    &
    \leq
    \frac{1}{3}
    \left(
      1 + \frac{2 B^{2}}{\lambda} T^{1/3}
    \right)
    \ln \frac{N}{\delta}
    +
    \frac{\sqrt{3} B}{\sqrt{\lambda}}
    T^{2/3} \sqrt{1 + \frac{4}{3 T}}
    \sqrt{\ln \frac{N}{\delta}}
    .
   \end{split}
  \end{equation}
\begin{proof}
As customary for concentration inequalities,
we start by a variance and size estimate:
\begin{equation*}
  \variance(t){\hat{L}_{t}^{\intercal} x - L_{t}^{\intercal} x}
  \leq
  \expectation(t){(\hat{L}_{t}^{\intercal} x)^{2}}
  \leq
  \expectation(t){x^{\intercal} C_{t}^{-1} x_{t} x_{t}^{\intercal}
    C_{t}^{-1} x}
  =
  x^{\intercal} C_{t}^{-1} x
  \leq
  \frac{B^{2}}{\gamma_{t} \lambda}
  ,
\end{equation*}
and
\begin{equation*}
  \abs{\hat{L}_{t}^{\intercal} x - L_{t}^{\intercal} x}
  \leq
  1
  +
  \frac{B^{2}}{\gamma_{t} \lambda}
  .
\end{equation*}
Also note that \(\hat{L}_{t}^{\intercal} x - L_{t}^{\intercal} x\)
is a martingale difference sequence.
Hence by Benett's inequality  (see Theorem~\ref{thm:Benett})
the claim follows
for a fixed action \(x\)
with probability at least \(1 - \delta / N\).
Therefore by the union bound,
it holds for all \(x \in A\) simultaneously
with probability at least \(1 - \delta\).
\end{proof}
\end{lemma}

Summing up \eqref{eq:EXP-simplified-Comb},
\eqref{eq:estimator-action-Comb}
(substituting \(\delta / (N+2)\) for \(\delta\)),
and
\eqref{eq:estimator-baseline-Comb}
(substituting \(N \delta / (N+2)\) for \(\delta\)),
we obtain \eqref{eq:ComBandit}
with probability at least \(1 - \delta\).

\section{Concluding remarks}
\label{sec:concluding-remarks}

We would like to mention that our method could be immediately strengthened
to provide an \emph{optimal}
high-probability regret of \(O(\sqrt{T})\) using the correction term
of \Algorithm{GeometricHedge} (see \cite{RegretLinearBandit2008}) and the identity 
\[\expectation(n){\sum_{i \in [d]} \tilde{M}_{i}(n) \tilde{X}_{i}^{2}(n)}
= \expectation(n){X(n)^{\intercal} M(n) M(n)^{\intercal}
  \Sigma_{n-1}^{+} \tilde{M}(n) \tilde{M}(n)^{\intercal}
  \Sigma_{n-1}^{+}  M(n) M(n)^{\intercal}X(n)},
\]
used for establishing the \(O(\sqrt{T})\)
regret bound for the expected case under oblivious adversaries in
\cite[supplementary material, proof of Theorem 6]{CombEXP2015}.
However, we were unable to verify this identity
\footnote{As of October 2016, we are discussing the matter with the
  authors of \cite{CombEXP2015}.},
which is equivalent
to 
\[\expectation(n){\sum_{i \in [d]} \tilde{M}_{i}(n) \tilde{X}_{i}(n)^{2}}
= \expectation(n){\left( \sum_{i \in [d]} \tilde{M}_{i}(n)
    \tilde{X}_{i}(n) \right)^{2}},\]
and as such we only claim the weaker bound of
\(O(T^{2/3})\).
This is the \emph{only} obstacle to combining \Algorithm{CombEXP} with
\Algorithm{GeometricHedge} to obtain an efficient algorithm with
optimal high-probability regret \(O(\sqrt{T})\) for the adaptive case
using our method. 

To put this into context, without the above identity also for the expected regret
case under oblivious adversaries we were only able
to establish an \(O(T^{2/3})\) regret bound, matching our high-probability
regret bound for adaptive adversaries.

\bibliographystyle{abbrvnat}
\bibliography{bibliography}

\newpage

\appendix

\section{Time-varying \Algorithm{EXP} algorithm 
  with projections}
\label{sec:time-exp3-proj}

Let \(\R^n_{>0} \coloneqq \set{x \in \R^{n}_{+}}{x_{i} > 0
  \text{ for all } i \in [n]}\)
be the \emph{strictly positive orthant}.
In this section we provide a version of \Algorithm{EXP}
(see Algorithm~\ref{alg:EXP3})
\begin{enumerate*}
\item that computes points in an arbitrary convex set
  \(P \subseteq \R^n_{>0}\)
  (as compared to distributions in the probability simplex)
\item that is \emph{any-time}, i.e., the parameter choice is
independent of \(T\) and the regret bounds hold uniformly for any \(t
\leq T\)
\end{enumerate*}
We explicitly allow arbitrary
dependence between the parameters \(\eta_{t}\),
input \(L_{t}\), and the points \(x_{t}\)
computed by the algorithm, to ease the use of the regret bound
in applications.

\begin{algorithm}
  \caption{\Algorithm{EXP} for convex sets contained in \(\R_{>0}^n\) with time varying
    parameters}
  \label{alg:EXP3}
  \begin{algorithmic}
    \REQUIRE convex set \(P \subseteq \R^{n}_{>0}\),
    start point \(x_{1} \in P\),
    loss vectors \(L_{t} \in \R^{n}\),
    positive parameters \(\eta_{1} \geq \eta_{2} \geq \dots\),
      satisfying \(\eta_{t} L_{t, i} \geq -1\)
      for \(i=1, \dotsc, n\)
    \FOR{\(t=1\) \TO \(T - 1\)}
      \STATE \(\tilde{x}_{t+1, i} \leftarrow
      x_{1,i}^{1 - \eta_{t+1} / \eta_{t}}
      x_{t, i}^{\eta_{t+1} / \eta_{t}} \cdot
      e^{- \eta_{t+1} L_{t, i}} \ \text{ for all } i \in [n]\) 
      \STATE
        Find \(x_{t+1} \in P\) such that
        \(\KLdivergence(z, x_{t+1})
        \leq \KLdivergence(z, \tilde{x}_{t+1}) + \varepsilon_{t}\)
        for all \(z \in P\).
        \COMMENT{approximate Bregman projection}
    \ENDFOR
  \end{algorithmic}
\end{algorithm}

\begin{lemma}
  \label{lem:EXP3}
  Let \(P \subseteq \R^{n}_{>0}\) be a convex set and let \(\eta_{t} L_{t, i} \geq -1\)
  for all \(t\) and \(i\),  the vector \(x_{t}\) computed by Algorithm~\ref{alg:EXP3}
  satisfy the following:
  \begin{equation}
    \label{eq:EXP3}
    \sum_{t=1}^{T} L_{t}^{\intercal} x_{t}
    -
    \sum_{t=1}^{T} L_{t}^{\intercal} y
    \leq
    \frac{\KLdivergence(y, x_{1})}{\eta_{T}}
    +
    \sum_{t=1}^{T-1} \frac{\varepsilon_{t}}{\eta_{t+1}}
    +
    (e - 2)
    \sum_{t=1}^{T} \eta_{t} \sum_{i=1}^{n} x_{t, i} L_{t, i}^{2}
    .
  \end{equation}
\begin{proof}
The proof is an extension of the standard analysis of
\Algorithm{EXP}, using the potential
\((\KLdivergence(y, x_{t}) - \KLdivergence(y, x_{1})) / \eta_{t}\)
to measure progress:
\begin{equation}
  \label{eq:EXP3-potential}
  \KLdivergence(y, x_{t}) - \KLdivergence(y, x_{1})
  =
  \sum_{i=1}^{n} y_{i} \ln \frac{x_{1, i}}{x_{t, i}}
  -
  \sum_{i=1}^{n} x_{1, i}
  +
  \sum_{i=1}^{n} x_{t, i}
  .
\end{equation}
We compare this with the potential in the next round,
first using \(\tilde{x}_{t+1}\) instead of \(x_{t+1}\):
\begin{multline*}
  \KLdivergence(y, \tilde{x}_{t+1}) - \KLdivergence(y, x_{1})
  =
  \frac{\eta_{t+1}}{\eta_{t}}
  \sum_{i=1}^{n} y_{i} \ln \frac{x_{1, i}}{x_{t, i}}
  +
  \eta_{t+1} \sum_{i=1}^{n} y_{i} L_{t, i}
  -
  \sum_{i=1}^{n} x_{1, i}
  +
  \sum_{i=1}^{n} x_{1, i}
  \left(
    \frac{x_{t, i}}{x_{1, i}} e^{- \eta_{t} L_{t, i}}
  \right)^{\eta_{t+1} / \eta_{t}}
  \\
  \leq
  \frac{\eta_{t+1}}{\eta_{t}}
  \sum_{i=1}^{n} y_{i} \ln \frac{x_{1, i}}{x_{t, i}}
  +
  \eta_{t+1} L_{t}^{\intercal} y
  -
  \sum_{i=1}^{n} x_{1, i}
  +
  \sum_{i=1}^{n} x_{1, i}
  \left[
    1 + \frac{\eta_{t+1}}{\eta_{t}}
    \left(
      \frac{x_{t, i}}{x_{1, i}} e^{- \eta_{t} L_{t, i}}
      -
      1
    \right)
  \right]
  \\
  =
  \frac{\eta_{t+1}}{\eta_{t}}
  \left(
    \KLdivergence(y, x_{t}) - \KLdivergence(y, x_{1})
    +
    \sum_{i=1}^{n} x_{t, i} (e^{- \eta_{t} L_{t, i}} - 1)
  \right)
  + \eta_{t+1} L_{t}^{\intercal} y
  ,
\end{multline*}
removing the exponent \(\eta_{t+1} / \eta_{t}\)
using \(z^{a} \leq 1 + a (z - 1)\) for \(z > 0\) and \(0 < a < 1\)
(which is Jensen's inequality for \(z^{a}\) as a function of \(a\)),
and then plugging in \eqref{eq:EXP3-potential}.
Rearranging and using the estimate
\(e^{a} \leq 1 + a + (e - 2) a^{2}\) for \(a \leq 1\)
with the choice \(a \coloneqq -\eta_{t} \hat{L}_{t, i} \leq 1\)
provides:
\begin{multline*}
  \frac{\KLdivergence(y, \tilde{x}_{t+1})
    - \KLdivergence(y, x_{1})}{\eta_{t+1}}
  -
  \frac{\KLdivergence(y, x_{t})
    - \KLdivergence(y, x_{1})}{\eta_{t}}
  - L_{t}^{\intercal} y
  \leq
  \frac{1}{\eta_{t}}
  \sum_{i=1}^{n}
  x_{t, i}
  \left(
    e^{- \eta_{t} L_{t, i}}
    -
    1
  \right)
  \\
  \leq
  \frac{1}{\eta_{t}}
  \sum_{i=1}^{n}
  x_{t, i}
  \left(
    - \eta_{t} L_{t, i}
    + (e - 2) (\eta_{t} L_{t, i})^{2}
  \right)
  =
  - L_{t}^{\intercal} x_{t}
  +
  (e - 2) \eta_{t} \sum_{i=1}^{n} x_{t, i} L_{t, i}^{2}
  ,
\end{multline*}
Summing up for \(t=1, \dots, T\) and rearranging leads to
(using the value \(\eta_{T+1} = \eta_{T}\))
\begin{equation*}
 \begin{split}
  \sum_{t=1}^{T} L_{t}^{\intercal} x_{t}
  - \sum_{t=1}^{T} L_{t}^{\intercal} y
  - (e - 2) \sum_{t=1}^{T} \eta_{t} \sum_{i=1}^{n} x_{t, i} L_{t, i}^{2}
  &
  \leq
  \sum_{t=1}^{T-1}
  \frac{\KLdivergence(y, x_{t+1})
    - \KLdivergence(y, \tilde{x}_{t+1})}{\eta_{t}}
  +
  \frac{\KLdivergence(y, x_{1})
    - \KLdivergence(y, \tilde{x}_{T+1})}{\eta_{T}}
  \\
  &
  \leq
  \sum_{t=1}^{T-1} \frac{\varepsilon_{t}}{\eta_{t+1}}
  +
  \frac{\KLdivergence(y, x_{1})}{\eta_{T}}
  ,
 \end{split}
\end{equation*}
using \(\KLdivergence(y, \tilde{x}_{T+1}) \geq 0\)
and
\(\KLdivergence(y, x_{t+1}) - \KLdivergence(y, \tilde{x}_{t+1})
\leq \varepsilon_{t}\).
The claim follows by rearranging.
\end{proof}
\end{lemma}

\section{Concentration inequalities}
\label{sec:concentration}

We will use the following concentration inequalities.

\begin{theorem}[Azuma–Hoeffding inequality]
  \label{thm:Azuma}
  For a martingale difference sequence \(X_{t}\)
  with \(a_{t} \leq X_{t} \leq b_{t}\)
  almost surely for constants \(a_{t}\), \(b_{t}\),
  we have with probability at least \(1 - \delta\)
  \begin{equation}
    \label{eq:Azuma}
    \sum_{t=1}^{T} X_{t} \leq
    \sqrt{\frac{\sum_{t=1}^{T} (b_{t} - a_{t})^{2}
        \ln (1 / \delta)}{2}}
    .
  \end{equation}
\end{theorem}

While the following inequality is stated only for \(b=1\)
in \cite[(18)]{HoeffdingSuper2012} it easily generalizes
via scaling to arbitrary \(b > 0\).
\begin{theorem}[{Benett's inequality \cite[(18)]{HoeffdingSuper2012}}]
  \label{thm:Benett}
  For a supermartingale difference sequence \(X_{t}\)
  bounded above by a positive constant \(X_{t} \leq b\),
  for any \(v \geq 0\)
  with probability at least \(1 - \delta\):
  \begin{align*}
  \sum_{t=1}^{T} \variance(t){X_{t}} &\geq v
    &&\text{or}
    &
    \sum_{t=1}^{T} X_{t}
    &
    \leq
    \frac{b \ln (1 / \delta)}{3}
    +
    \sqrt{2 v \ln (1 / \delta)}
    .
  \end{align*}
\end{theorem}

\section{Projection for Kullback–Leibler divergence}
\label{sec:projection-KL}

We will now describe
a generic, efficient, simple Frank–Wolfe algorithm
for the projection step in Line~\ref{line:projection}
of Algorithm~\ref{alg:CombEXP}.
We remark that there are many possibilities for improvements,
such as, e.g., employing advanced variants of the Frank–Wolfe algorithm
(see e.g., \cite{FW-converge2015}) or using customized algorithms for
specific polytopes. For example, in the case of the simplex \(P = \set{x \geq 0}{\sum_{i} x_{i} = 1}\),
the projection of \(x\) is simply \(x / \sum_{i=1}^{n} x_{i}\) and for
the the permutahedron there exist very fast, specialized projection
methods (see e.g., 
\cite{projectPermutahedron2016}).

\begin{algorithm}
  \caption{Projection for \(\KLdivergence\)}
  \label{alg:projection-KL}
  \begin{algorithmic}
    \REQUIRE
      linear optimization oracle over a polytope
      \(P \subseteq [\alpha, \beta]^{n}\),
      \(\alpha > 0\),
      upper bound \(B\) for the \(\ell_{2}\)-diameter of \(P\),
      accuracy \(\varepsilon > 0\),
      point \(x \in \R^{n}_{> 0}\)
    \ENSURE
      \(y_{K} \in P\) with
      \(\KLdivergence(z, y_{K}) \leq \KLdivergence(z, x)
      + \varepsilon\)
      for all \(z \in P\)
    \STATE \(y_{0} \in P\) any point
    \STATE
      \(K \leftarrow \left\lceil
        \frac{4 B^{4} \beta}{\alpha^{3} \varepsilon^{2}} \right\rceil\)
    \FOR{\(k=1\) \TO \(K\)}
      \STATE \(s \in \argmin_{z \in P}
        \sum_{i=1}^{n} z_{i} \ln (y_{k-1, i} / x_{i})\)
        \COMMENT{Linear optimization oracle call}
      \STATE \(y_{k} \leftarrow ((k-1) y_{k-1} + 2 s) / (k+1)\)
    \ENDFOR
    \RETURN \(y_{K}\)
  \end{algorithmic}
\end{algorithm}

\begin{proposition}
  \label{prop:projection-KL}  Given a polytope \(P \subseteq [\alpha,
  \beta]^{n}\) with \(\alpha > 0\),
  an upper bound \(B\) for the \(\ell_{2}\)-diameter
      of \(P\), as well as an accuracy \(\varepsilon > 0\),
  Algorithm~\ref{alg:projection-KL} computes an approximate
  projection with
  \(O \left( \frac{B^{4} \beta}{\alpha^{3} \varepsilon^{2}} \right)\)
  oracle calls.
\begin{proof}
As the algorithm calls the oracle once per iteration,
the bound on the number of oracle calls is immediate.
To prove the claimed accuracy of the returned point \(y_{K}\),
note that the algorithm is the Frank–Wolfe algorithm
for the function \(f(z) \coloneqq \KLdivergence(z, x)\).
Recall that the gradient \(\nabla f(z)\) of \(f\) at \(z\) is given by
\((\nabla f(z))_{i} = \ln (z_{i} / x_{i})\)
and the Hessian is a diagonal matrix
\(\nabla^{2} f(z) = \diag(1 / z_{1}, 1/z_{2}, \dotsc, 1/z_{n})\).
As \(1/\beta \leq 1/z_{i} \leq 1/\alpha\) for \(z \in P\),
the function \(f\) is \(1/\alpha\)-smooth
and \(1/\beta\)-strongly convex on \(P\) in the \(\ell_{2}\)-norm,
and has curvature \(C_{f} \leq B^{2} / \alpha\).
Let \(x^{*} \coloneqq \argmin_{z \in P} f(z)\),
i.e., the Bregman projection of \(x\) to \(P\).
By \cite[Theorem~1]{jaggi2013revisiting},
\(f(y_{K}) - f(x^{*}) \leq 2 C_{f} / (K + 2)\),
therefore by strong convexity
\begin{equation}
  \frac{1}{2 \beta} \norm[2]{y_{K} - x^{*}}^{2}
  \leq
  \KLdivergence(y_{K}, x) - \KLdivergence(x^{*}, x)
  \leq \frac{2 B^{2}}{\alpha (K + 2)}
  .
\end{equation}
Let \(z \in P\) be arbitrary.  By the Pythagorean Theorem we have
\(\KLdivergence(z, x^{*}) \leq \KLdivergence(z, x)\) and thus
\begin{equation*}
 \begin{split}
  \KLdivergence(z, y_{K}) - \KLdivergence(z, x)
  &
  \leq
  \KLdivergence(z, y_{K}) - \KLdivergence(z, x^{*})
  =
  \sum_{i=1}^{n} z_{i} \ln \frac{x^{*}_{i}}{y_{K, i}}
  -
  \sum_{i=1}^{n} x^{*}_{i} + \sum_{i=1}^{n} y_{K, i}
  \\
  &
  \leq
  \sum_{i=1}^{n} z_{i} \left( \frac{x^{*}_{i}}{y_{K, i}} - 1 \right)
  -
  \sum_{i=1}^{n} x^{*}_{i} + \sum_{i=1}^{n} y_{K, i}
  =
  \sum_{i=1}^{n}
  \frac{z_{i} - y_{K, i}}{y_{K, i}} (x^{*}_{i} - y_{K, i})
  \leq
  \frac{B}{\alpha}
  \norm[2]{x^{*} - y_{K}}
  \\
  &
  \leq
  \frac{B \sqrt{2 \beta}}{\alpha}
  \sqrt{\KLdivergence(y_{K}, x) - \KLdivergence(x^{*}, x)}
  \leq
  \frac{2 B^{2} \sqrt{\beta}}{\alpha^{3/2} \sqrt{K + 2}}
  \leq \varepsilon
  .
 \end{split}
\end{equation*}
Plugging in \(K = \left\lceil
        \frac{4 B^{4} \beta}{\alpha^{3} \varepsilon^{2}} \right\rceil\) as set by the algorithm provides the result.
\end{proof}
\end{proposition}

\section{Linear decomposition}
\label{sec:linear-decomposition}

For the convenience of the reader,
we briefly recall the decomposition algorithm
(Algorithm~\ref{alg:linear-decomposition}) of \cite{Caratheodory2015}
that for a polytope \(P\)
approximately decomposes any point \(x \in P\)
into a convex combination of vertices of \(P\),
using a linear optimization oracle over \(P\).
The algorithm uses Mirror Descent
(see \cite{nemirovski1979efficient})
to find a convex combination.

\begin{proposition}[{\cite[Theorem 3.5]{Caratheodory2015}}]
  \label{prop:linear-decomposition}
  Given a polytope \(P\)
  with diameter at most \(2 D\) in \(\ell_{2}\)-norm,
  and a point \(x \in P\),
  Algorithm~\ref{alg:linear-decomposition}
  computes with \(O(D^{2} / \varepsilon^{2})\)
  calls to a linear optimization oracle
  over \(P\)
  a multiset \(x_{1}\), \dots, \(x_{k}\) of vertices
  for \(k = \lceil 4 D^{2}/\varepsilon^{2} \rceil\)
  such that \(\norm[2]{\sum_{i=1}^{k} x_{i}/k - x} \leq \varepsilon\).
\end{proposition}

\begin{algorithm}
  \caption{Linear decomposition}
  \label{alg:linear-decomposition}
  \begin{algorithmic}
    \REQUIRE linear optimization oracle over polytope \(P\),
      an inner point \(x \in P\), precision \(\varepsilon\)
    \ENSURE vertices \(x_{1}, \dots, x_{k} \in P\) such that
      \(\norm[2]{x - \sum_{i} \lambda_{i} x_{i} / k} \leq \varepsilon\)
    \STATE \(k \leftarrow \lceil 4 D^{2} / \varepsilon^{2} \rceil\)
    \STATE \(\eta \leftarrow 4 \varepsilon (p - 1)\)
    \STATE \(y_{1} \leftarrow 0\); \(z_{1} \leftarrow 0\)
    \FOR{\(t = 1\) \TO \(k\)}
      \STATE Choose vertex
      \(x_{t} \in \argmin_{y \in P} y_{t}^{\intercal} y\)
      \COMMENT{Linear optimization oracle call}
      \STATE \(z_{t+1} \leftarrow z_{t} - \eta (x - x_{t})\)
      \IF{\(\norm[2]{z_{t+1}} > 1\)}
        \STATE \(y_{t+1} \leftarrow z_{t+1} / \norm[2]{z_{t+1}}\)
      \ELSE
        \STATE \(y_{t+1} \leftarrow z_{t+1}\)
      \ENDIF
    \ENDFOR
    \RETURN \(x_{1}\), \dots, \(x_{k}\)
  \end{algorithmic}
\end{algorithm}

\section{Fitness of barycentric spanners for exploration}
\label{sec:fitness-spanner}

Let \(\lambda_{\min}(\mu)\) denote the minimal eigenvalue of the
covariance matrix \(\expectation(x \sim \mu){x x^{\intercal}}\)
of a distribution \(\mu\).
For exploration one wishes to find a \(\mu\) with
a high minimal eigenvalue \(\lambda_{\min}(\mu)\).
Here we show that a uniform distribution on
any approximate barycentric spanner achieves within an \(O(n^{2})\)
factor the best possible minimal eigenvalue using any scalar product
on \(\R^{n}\).
The free choice of scalar product and hence orthonormal basis
allows preserving sparse representation of a polytope \(P\).

\begin{lemma}
  \label{lem:fitness-spanner}
  Let \(v_{1}, \dotsc, v_{n}\) be a
  \(C\)-approximate barycentric spanner of a polytope
  \(P \subseteq \R^{n}\).
  Then the uniform distribution \(\mu_{v_{1}, \dotsc, v_{n}}\)
  on the spanner satisfies
  \begin{equation}
    \label{eq:fitness-spanner}
    \lambda_{\min}(\mu_{v_{1}, \dotsc, v_{n}})
    \geq
    \frac{\lambda_{\min}(\mu)}{C^{2} n^{2}}
  \end{equation}
  for any distribution \(\mu\) over \(P\).
\begin{proof}
Using that the \(v_{i}\) form a barycentric spanner,
there are coefficients \(\lambda_{x, i}\) for all \(x \in P\)
satisfying
\begin{equation*}
  x = \sum_{i} \lambda_{x, i} v_{i},
  \qquad
  \abs{\lambda_{x, i}} \leq C
  .
\end{equation*}
In particular,
with
\(\alpha_{x} \coloneqq \sum_{i=1}^{n} \abs{\lambda_{x, i}} \leq C n\)
by Jensen's inequality
\begin{equation*}
  x x^{\intercal}
  \preceq
  \sum_{i=1}^{n} \alpha_{x} \abs{\lambda_{x, i}} v_{i} v_{i}^{\intercal}
  \preceq
  C^{2} n \sum_{i=1}^{n} v_{i} v_{i}^{\intercal}
  .
\end{equation*}
Hence
\(\expectation(x \sim \mu){x x^{\intercal}}
  \preceq
  C^{2} n \sum_{i=1}^{n} v_{i} v_{i}^{\intercal}
  =
  C^{2} n^{2} \expectation(x \sim \mu_{v_{1}, \dotsc, v_{n}}){x
    x^{\intercal}}\),
from which the claim follows.
\end{proof}
\end{lemma}

\end{document}